\documentclass[aps,prl,twocolumn,10pt,superscriptaddress]{revtex4-1}
\usepackage[pdftex]{color,graphicx}
\usepackage{array}
\usepackage{amsmath}
\usepackage{amssymb}

\begin{document}

\title{Semiconducting nanotubes derived from a rectangular graphyne: a DFT study}

\author{Wjefferson Henrique da Silva Brandão}
 \affiliation{Institute of Physics, Fluminense Federal University, Niterói, 24210-340, Rio de Janeiro, Brazil.}
 
\author{Anderson Gomes Vieira}
 \affiliation{Departamento de F\'isica, Universidade Federal do Piau\'i, Teresina, 64049-550, Piau\'i, Brazil} 
 
\author{Jonathan da Rocha Martins}
 \affiliation{Departamento de F\'isica, Universidade Federal do Piau\'i, Teresina, 64049-550, Piau\'i, Brazil}

 \author{Andrea Latgé}
 \affiliation{Institute of Physics, Fluminense Federal University, Niterói, 24210-340, Rio de Janeiro, Brazil.}

\author{Marcelo Lopes Pereira Junior}
 \affiliation{University of Brasília, College of Technology, Department of Electrical Engineering, Brasília, 70910-900, Brazil.}
 
\author{Eduardo Costa Gir\~ao}
 \affiliation{Departamento de F\'isica, Universidade Federal do Piau\'i, Teresina, 64049-550, Piau\'i, Brazil}

 \date{\today}
\begin{abstract}
Proposing new ways to organize carbon in 2D nanomaterials has been a relevant strategy in the search for systems with targeted properties for different applications. One focus is the study of fully sp$^2$ non-graphitic networks, with successfully synthesized examples. Hybrid sp-sp$^2$ systems of the graphyne family are a related approach, and many systems have the honeycomb lattice as a base model. However, other examples have been inspired by other lattices as the recently proposed r$\gamma$GY sheet, which features a semiconducting behavior with highly localized \emph{quasi}-1D states. Here, we investigate how to tune r$\gamma$GY properties by folding this sheet into nanotube forms. We elucidate mechanisms that determine their electronic structure by means of density functional theory calculations, as well as we identify the interplay involving chirality, diameter, and the emergence of dispersive/localized frontier states on gap modulation through simple extrapolated methods.
\end{abstract}

\maketitle

\section{Introduction}

Fullerenes are usually cast as the starting point of the nanocarbon materials timeline~\cite{kroto1985}. However, carbon nanotubes were the driving force behind the development of a worldwide scientific community, leading to the broad field of nanocarbons as we know it today~\cite{iijima1991,iijima1993,bethune1993,hughes2024}. While sharing space with graphene~\cite{novoselov2004,zhao2024} and other 2D materials~\cite{manzeli2017,muzaffar2021} over the last two decades, nanotubes have remained promising for several applications~\cite{hughes2024}. They exhibit selectivity to interaction with specific compounds, making them potentially useful for sensors and extraction methods~\cite{onyancha2021}, as well as their electronic properties dependent on curvature and boundary conditions in an intricate way~\cite{Latge2000,dai2002,samsonidze2003}. Their geometry is naturally convenient for nanoelectronic applications as both electrodes~\cite{qi2004} and active parts of nanodevice setups~\cite{kleinherbers2023}. Given their relevance, as exemplified by the topic listed above, the relation between the properties of nanotubes and the graphene sheet that makes up the tube walls has emerged as a basic topic in carbon-based textbooks~\cite{saito1998}.
%and in many disciplines for graduate courses~\cite{saito1998}.
 
Graphene has emerged as one of the most extensively studied nanocarbon materials since the groundbreaking experiments of the early 2000s~\cite{novoselov2004}. Considerable research efforts have focused on tailoring its electronic properties, particularly in strategies aimed at opening a band gap~\cite{terrones2012,lee2018}. In this context, although carbon nanotubes have been systematically investigated for a longer period, they represent a natural extension of the graphene framework for tuning the physical behavior of the honeycomb lattice. Unlike nanoribbons~\cite{son2006b}, whose properties are strongly influenced by edge effects ~\cite{son2006,pisani2007}, carbon nanotubes preserve the structural integrity of the parent lattice, facilitating direct comparison with the properties of the planar graphene sheet.

Efforts to tailor the electronic properties of nanocarbon materials, especially graphene, date back to the late 1990s, when the idea of engineering new two-dimensional carbon lattices was first proposed~\cite{crespi1996}.
%When it comes to adjusting the electronic behavior of nanocarbon (graphene, in particular), a hypothetical pathway with grounds in the late 1990s is the proposal of new carbon lattices in two dimensions~\cite{crespi1996}. 
Early ideas on this issue were mainly related to the distribution of extended sets of defects in graphene until a point is reached at which the structure cannot be viewed as defective graphene, but as a new material~\cite{terrones2000}. These were the Haeckelite structures, which can contain pentagonal and heptagonal rings in addition to the original hexagons of graphene~\cite{liu2021,chen2024}. In the following decades, several other proposals for different nanocarbon 2D lattices have been reported, covering a broad range of behaviors~\cite{girao2023}. Such studies gained additional motivation, as a non-graphitic lattice was synthesized a few years ago~\cite{fan2021}. As the field evolved, such proposals began to involve not only the sp$^2$ state of carbon, but also atoms with other hybridization states~\cite{zhang2015,ram2018,malko2012}. A remarkable example is the class of graphynes, carbon sheets composed of a combination of sp and sp$^2$ carbons~\cite{ivanovskii2013}. They were initially hypothesized from the insertion of acetylenic bridges in substitution of sp$^2$-sp$^2$ carbon bonds in graphene, so that different concentrations and different strategies to insert the sp chains result in several distinct graphynes with electronic properties ranging from metallic to Dirac materials and semiconductors~\cite{malko2012,wu2013}. In analogy to graphene, the electronic behavior of graphynes can be further tuned by their modification into nanoribbon and nanotube forms~\cite{yue2012,kang2015,mohammadi2018,Latge2022}. 

Finally, a more recent trend has been the study of graphyne forms not inspired by graphene, but in other sp$^2$ 2D allotropes of carbon~\cite{oliveira2022,ullah2024,rego2025}. One recent example is a system called rectangular $\gamma$-graphene (r$\gamma$GY), a graphyne structure that does not have graphene, but 2D biphenylene as its parent sp$^2$ sheet~\cite{vieira2024}. The symbol  $\gamma$ in r$\gamma$GY refers to the fact that it preserves a single hexagonal sp$^2$ inside its unit cell, with subsequent hexagons connected by acetylenic connections.

The r$\gamma$GY sheet has been reported as a semiconducting sheet with anisotropic electronic properties, as highly localized states extend along a single crystalline direction of the sheet. In contrast, its frontier states are shown to be adjusted by the size of the acetylenic chains composing the system~\cite{vieira2024}. Later, r$\gamma$GY was further studied with a class of other graphyne systems~\cite{hong2024}. Here, we follow the analogy with the graphene case and study the electronic properties of r$\gamma$GY nanotubes (r$\gamma$GYNTs). We show how the tube properties vary with different chiralities, as well as how they are affected by curvature effects. These tubes are generally semiconductors, but with substantial differences from their 2D counterparts, especially in high-curvature cases. The results are further rationalized by a zone-folding approach that reveals the origins of flat bands present in certain kinds of tubes.

\section{Studied systems}

\begin{figure}[t!]
\includegraphics[width=\columnwidth]{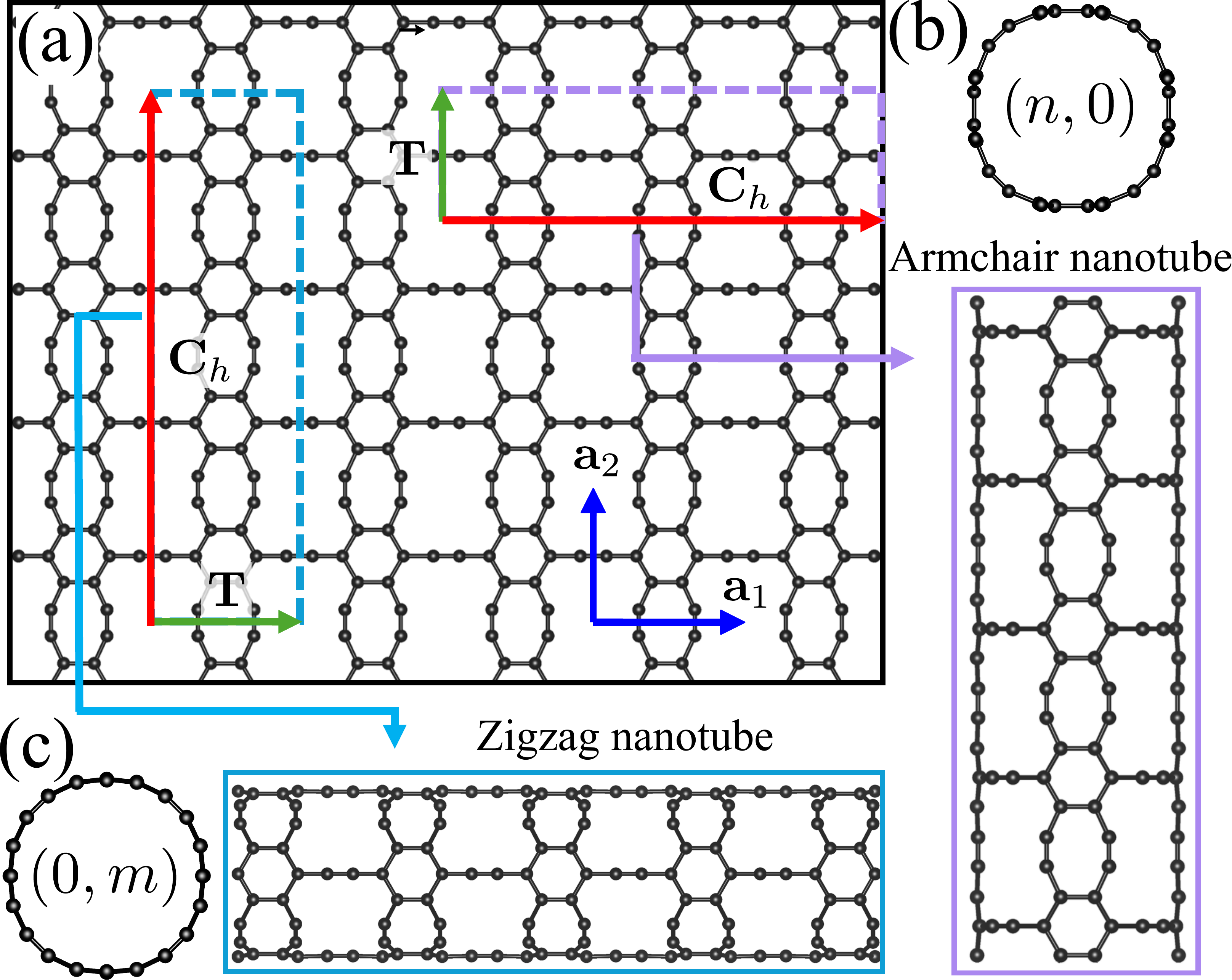}
\caption{(a) Atomic structure of the r$\gamma$GY sheet, with its lattice vectors $\mathbf{a}_1$ and $\mathbf{a}_2$ represented by the blue arrows. Here, the projections of the unit cells of the (3,0) and (0,4) tubes over the r$\gamma$GY sheet are represented by the dashed purple and blue rectangles, while their $\mathbf{C}_h$ ($\mathbf{T}$) vectors are represented by red (green arrows). (b) Example of a $(n,0)$ nanotube with $n=4$. (c) Example of a $(0,m)$ nanotube with $m=4$.}
\label{fig-systems}
\end{figure}

In Fig.~\ref{fig-systems}(a) we illustrate the r$\gamma$GY sheet, together with its lattice vectors $\mathbf{a}_1=(a,0)$ and $\mathbf{a}_2=(0,b)$, where the optimized lattice parameters are $a=6.92$~\AA~and $b=6.28$~\AA. Considering the original $sp^2$ 2D biphenylene as the starting point, acetylenic bridges are introduced on bonds linking different hexagons along both the $\mathbf{a}_1$ and $\mathbf{a}_2$ directions. The acetylenic bridges along the $\mathbf{a}_2$ direction are not perfectly linear, and will hereafter be referred to as \emph{curved acetylenic bridges} (or CABs) from r$\gamma$GY. On the other hand, the chains connecting hexagons along the $\mathbf{a}_1$ direction are perfectly linear and will be referred to as \emph{straight acetylenic bridges} (or simply SABs). 

We studied nanotubes based on the r$\gamma$GY lattice, where the chiral vector defines the folding direction
\begin{equation}
\mathbf{C}_h=n\mathbf{a}_1+m\mathbf{a}_2=(n,m)\text{,}
\end{equation}
where $n$ and $m$ are integers, in analogy to the well-known case of nanotubes based on graphene. However, due to the rectangular symmetry of r$\gamma$GY, the range of $(n,m)$ values resulting in distinct nanotube geometries corresponds to $n,m\ge0$, as far as at least one of the chiral indexes $n$ and $m$ is nonzero. The projection of the nanotube unit-cell over the r$\gamma$GY sheet is further defined with the aid of a translational vector
\begin{equation}
\mathbf{T}=t_1\mathbf{a}_1+t_2\mathbf{a}_2=(t_1,t_2),
\end{equation}0
which is also defined in terms of integer $t_1$ and $t_2$ coefficients, but in such a way that $\mathbf{C}_h\cdot\mathbf{T}=0$. It turns out that $t_1$ and $t_2$ usually have very high values due to the non-integer ratio between the $\mathbf{a}_1$ and $\mathbf{a}_2$ lattices of r$\gamma$GY, which makes simulations impractical for arbitrary chiralities. A marked exception corresponds to the high-symmetric $(n,0)$ and $(0,m)$ cases, as their translational vectors are $(0,1)$ and $(1,0)$, respectively. For these reasons, we focus our study on $(n,0)$ and $(0,m)$ nanotubes with $n$ and $m$ ranging from 2 to 12. In Fig.~\ref{fig-systems}(b-c), we illustrate examples of a $(n,0)$ and a $(0,m)$ tube, while the projection of their unit cells over the r$\gamma$GY sheet is represented in Fig.~\ref{fig-systems}(a). The $(2,0)$ and $(0,2)$ cases (here considered limiting nanotube structures) feature very small $\approx$4~\AA~diameters, the reason for which we exclude hypothetical $(1,0)$ and $(0,1)$, since their diameters would be of the same order as a carbon-carbon bond. In Fig.~\ref{fig-systems}(b-c), for examples of the $(n,0)$ and $(0,m)$ nanotubes, we represent the projection of the nanotube's unit cell over 2D r$\gamma$GY, as well as their cylindrical structures. In analogy to the graphene's case, we call the $(n,0)$ and $(0,m)$ structures as \emph{armchair} and \emph{zigzag} r$\gamma$GYNTs, a-r$\gamma$GYNTs and z-r$\gamma$GYNTs, respectively. These designations follow the alignment of the hexagonal rings of r$\gamma$GY relative to the nanotube's cross-section.

\section{Methods}

To study the 2D and 1D r$\gamma$GY systems, we performed simulations based on density functional theory (DFT)~\cite{hohenberg1964,kohn1965} using the SIESTA code~\cite{Soler2002}. A double-$\zeta$ polarized (DZP) basis set of numerical atomic orbitals~\cite{Soler2002} was used to expand the wavefunctions for valence electrons, whereas core electrons were treated through Troullier-Martins pseudopotentials~\cite{troullier1991}. The reach of the basis set functions was defined in SIESTA by the energy shift parameter, set as 0.05 eV~\cite{anglada2002}. Exchange-correlation energy has been considered by means of the generalized gradient approximation (GGA) according to the Perdew-Burke-Ernzerhof (PBE)  parameterization~\cite{perdew1996}. The real-space grid used to describe charge density was defined in terms of a 400 Ry mesh cutoff, while reciprocal space integrations were performed with a $40\times44\times1$ ($1\times1\times44$) [$1\times1\times40$] Monkhorst-Pack sampling for 2D r$\gamma$GY (armchair r$\gamma$GYNTs) [zigzag r$\gamma$GYNTs]. In these periodic-boundary-conditions calculations, vacuum regions of 40 \AA\ were included for all the non-periodic directions of the 1D and 2D systems to avoid interactions with their periodic images. Structural relaxation was performed for the atomic coordinates by considering a maximum force tolerance of 0.01~eV/\AA, and the unit cell vectors were optimized within a maximum tolerance of 0.1 GPa for stress components.

\section{Results}
\subsection{Energetic Stability}
\begin{figure}[t!]
    \centering
    \includegraphics[width=\linewidth]{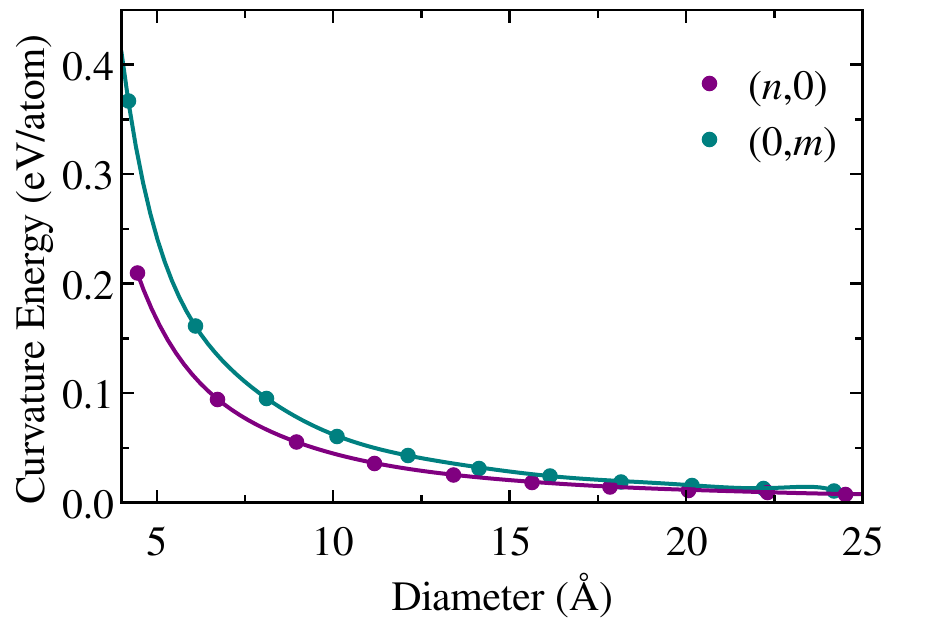}
    \caption{Curvature energy as a function of diameter for $(n,0)$ and $(0,m)$ r$\gamma$GYTs.}
    \label{fig:curv-energ}
\end{figure}

We analyze the structural stability of a nanotube relative to its 2D parent allotrope considering the curvature energy ($E_C$), which is the variation in total energy when the sheet is rolled to form the nanotube. For this analysis, shown in Fig. \ref{fig:curv-energ}, we define the curvature energy as being
\begin{equation}
    E_C=\dfrac{E_{1D}}{N_{1D}}-\dfrac{E_{2D}}{N_{2D}}
\end{equation}
where $E_{1D}$ and $E_{2D}$ ($N_{1D}$ and $N_{2D}$) are the total energies (number of atoms) in the 1D and 2D systems, respectively. Fig. \ref{fig:curv-energ} shows $E_C$ as a function of diameter for the two chiralities of the studied nanotubes.

Considering small diameters, the ($n$,0) family reveals a lower curvature energy compared to the (0,$m$) family, indicating that rolling the membrane in the $x$ direction requires less energy than rolling the sheet in the $y$ direction (see Fig. \ref{fig-systems}(a)). This preferential energetic direction can be explained by the structural and electronic anisotropy of the 2D sheet, which affects its rigidity to curve in different directions. In other words, the energy required to roll up the material depends on the crystallographic direction chosen, and this is directly related to the way the orbitals and carbon bonds are arranged in the plane. In particular, curvature in $(0,m)$ tubes affects mostly CABs, highlighting the locally less stable environment of these sp atoms, which deviate significantly from their ideal linear geometry. For larger diameters, $E_C$ tends to zero, acquiring values lower than $k_BT$ at room temperature already for the wider tubes considered in this work. This further indicates that the membrane is the ground state for the r$\gamma$GY lattice and that the energy cost of rolling the sheet into a nanotube becomes negligible as the radius increases. This behavior arises from the fact that, for small curvatures, the energy variation approximately follows a quadratic law $E_C\propto1/R^2$, where $R=d_t/2$ is the nanotube radius, and the energy difference between the flat and cylindrical configurations disappears when $R\rightarrow\infty$. In summary, the lower bending energy for the ($n$,0) family results from the mechanical anisotropy of the membrane, indicating that the $x$ direction has a lower curvature modulus than the $y$ direction. This indicates that the crystal lattice, its angular distribution of bonds, and electronic hybridization favor smoother deformations when the rolling occurs in this orientation.

\subsection{Electronic structure of nanotubes}

\begin{figure*}[ht!]
\includegraphics[width=\textwidth]{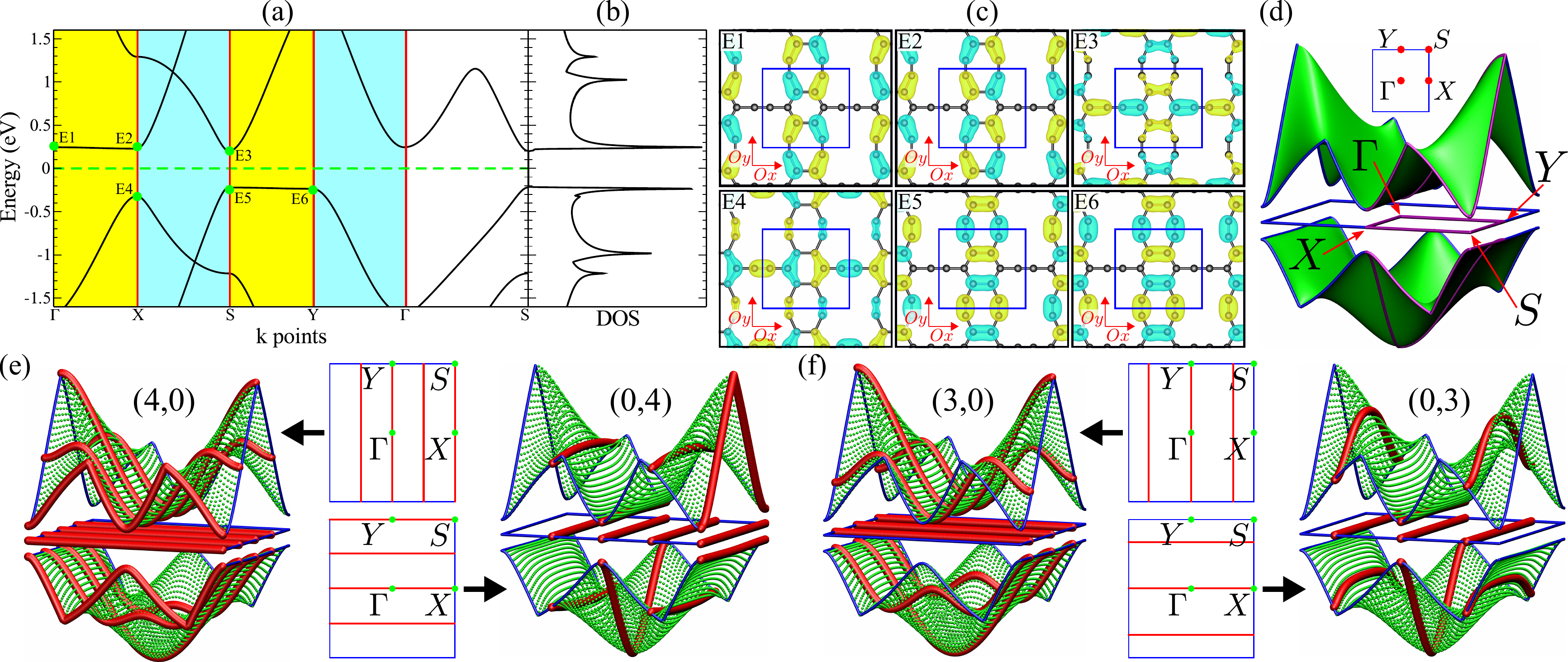}
\caption{(a) Electronic structure of r$\gamma$GY represented over high-symmetry lines of the BZ. The yellow (cyan) shaded areas represent the $k$-space paths parallel to the $\mathbf{b}_1$ ($\mathbf{b}_2$) vector of the reciprocal lattice. E1-E6 represent important energy levels from the frontier bands. (b) Density of states of r$\gamma$GY. (c) Real part of the wavefunction for the E1-E6 energy levels indicated in (a), with cyan and yellow clouds representing portions of the wavefunction with opposite sign. (d) Valence and conduction bands of r$\gamma$GY are represented by a surface plot over the entire BZ, with the Fermi energy represented by the blue rectangle. (e) Cutting lines for the (4,0) and (0,4) tubes represented over the 2D BZ and over the frontier bands of r$\gamma$GY. (f) Same as (e), but for the (3,0) and (0,3) tubes.}
\label{fig-rggy-ee}
\end{figure*}

Before discussing the electronic structure of nanotubes, it is essential to revisit the electronic structure of the r$\gamma$GY sheet. In Fig.~\ref{fig-rggy-ee}(a), we illustrate its electronic band structure represented along the high-symmetry lines of the rectangular Brillouin zone (BZ), followed by the corresponding density of states (DOS) in Fig.~\ref{fig-rggy-ee}(b). The valence band exhibits a flat sector along the $S-Y$ path, as well as a dispersive sector with a local maximum value at the $X$ point, which has a value slightly below that of the flat sector. The flat sector results in a prominent peak in the DOS plot at the valence band maximum (VBM) position, which is followed by a shoulder-shaped peak in the energy value of the local maximum at the $X$ point. A similar picture is observed for the conduction band, but with a flat sector along the $\Gamma-X$ path. Further, the dispersive sector of the conduction band has a minimum value at the $S$ point of the BZ, which is the global minimum of the band, but with an energy very close to that of the flat sector. The flat sector also results in a pronounced van Hove singularity at the conduction band minimum (CBM) (as visible from the DOS plot), and the global minimum of the valence band shows up as a small shoulder close below the CBM peak. All these features are in agreement with the previous study on the r$\gamma$GY sheet~\cite{vieira2024}. 

There are a few energy levels of the valence and conduction bands that deserve special attention, namely those marked by the green circles labeled as E1, E2, E3, E4, E5, and E6 in Fig.~\ref{fig-rggy-ee}(a). Such points will be relevant to understanding the electronic properties of r$\gamma$GYNTs, as discussed later. In Fig.~\ref{fig-rggy-ee}(c), we plot the real part of the wave functions of these levels over the r$\gamma$GY sheet, where the cyan and yellow sectors of the plots represent opposite signs of the wave function. The E1 and E2 levels lie at the $\Gamma$ and $X$ points of the BZ, respectively, both on the flat sector of the conduction band. They have very similar spatial distributions over the bonds involving 1) a sp$^2$ atom from the hexagonal rings and 2) a sp atom from the CABs along the $\mathbf{a}_2$ direction. They basically differ by the phase difference between neighboring unit cells due to the Bloch phase $\exp(i\mathbf{k}\cdot\mathbf{R})$ (where $i$ is the imaginary unit, $\mathbf{k}$ the vector from the BZ, and $\mathbf{R}$ a lattice vector). While the E1 wave function (at $\Gamma$) has the same phase for all unit cells, the E2 wave function (at $X$) inverts its sign for neighbor cells along the $\mathbf{a}_1$ direction, but not along $\mathbf{a}_2$. These wave functions are very similar to the local DOS (LDOS) plots from r$\gamma$GY over the flat sector of the conduction band~\cite{vieira2024}. A similar finding is observed when we plot the real part of the wave function for the E5 (at $S$) and E6 (at $Y$) levels from the flat sector of the valence band. These levels also distribute over the 8-carbon rings from r$\gamma$GY, as E1 and E2, but on complementary sets of bonds relative to E1/E2. Again, while the spatial distributions of E5 and E6 are similar to each other, they differ in their phases as we move over neighboring unit cells of the system. An explicit characteristic of the E1, E2, E5, and E6 states (from the flat sectors of the frontier bands) is their negligible contribution over the SABs aligned with the $\mathbf{a}_1$ direction of r$\gamma$GY, as reported before~\cite{vieira2024}, while the other two curved acetylenic bridges in the unit cell carry a significant contribution from these states. On the other hand, the E3 and E4 states are extreme values of the dispersive sectors of the conduction and valence bands, respectively, and exhibit significantly different distributions compared to the states from the flat bands. Namely, both E3 and E4 spread over the straight acetylenic bridges along the $\mathbf{a}_1$ direction, but on complementary sets of bonds relative to each other. As a result, we can easily identify the origin of a given frontier state (either from a flat or dispersive band sector) by looking at details of its spatial distributions. Namely, states E1, E2, E5, and E6 spread mainly on the CABs from r$\gamma$GY (and on the neighboring sp$^2$ sites from the hexagonal rings), while E3 and E4 primarily spread on the SABs from r$\gamma$GY (and on the neighboring sp$^2$ sites from the hexagonal rings). To simplify later discussions, we will extend the reference to acetylenic bridges along the $\mathbf{a}_1$ ($\mathbf{a}_2$) direction of r$\gamma$GY as SABs (CABs) and also to the nanotube counterparts.

The reason for revisiting the r$\gamma$GY sheet is that many aspects of the electronic structure of nanotubes can be understood from the band structure of the parent sheet by a zone folding (ZF) description. The ZF approach basically consists of 1) defining cutting lines over the ZB of the 2D system, 2) computing the electronic bands of the sheet for the $k$-points along these lines, and 3) using the last result as an approximation for the tubes' electronic band structures. Curvature effects in ZF are only indirectly considered, as the length and spacing for the set of cutting lines are determined by specific boundary conditions for each $(n,m)$ tube. At the same time, the ZF band values are still extracted from the Hamiltonian of the planar parent structure. For this reason, such an approach usually fails to give reasonable results for high-curvature tubes. Even so, ZF allows us not only to rationalize the results for very wide tubes, but also to understand how the tube's properties evolve as the diameter increases from narrow tubes. 

In Fig.~\ref{fig-rggy-ee}(d), we show the valence and conduction bands of r$\gamma$GY over the entire BZ by a surface plot, where we easily identify the flat and dispersive sectors of these bands. For the $(n,m)$ nanotube based on the r$\gamma$GY sheet, the cutting lines over the reciprocal space are given by the vectors $\mathbf{k}$ of the form
\begin{equation}
\mathbf{k}=j\mathbf{K}_1+\lambda\mathbf{K}_2,
\end{equation}
where $\mathbf{K}_1$ and $\mathbf{K}_2$ are the reciprocal lattice vectors corresponding to $\mathbf{C}_h$ and $\mathbf{T}$ from direct space, respectively. The integer $j$ index varies from 0 to $N-1$ ($N$ being the number of unit cells of the 2D system that fit within the unit cell of the nanotube) and $-1/2<\lambda\le1/2$~\cite{samsonidze2003,silva2019}. 

\begin{figure*}[ht!]
\includegraphics[width=\textwidth]{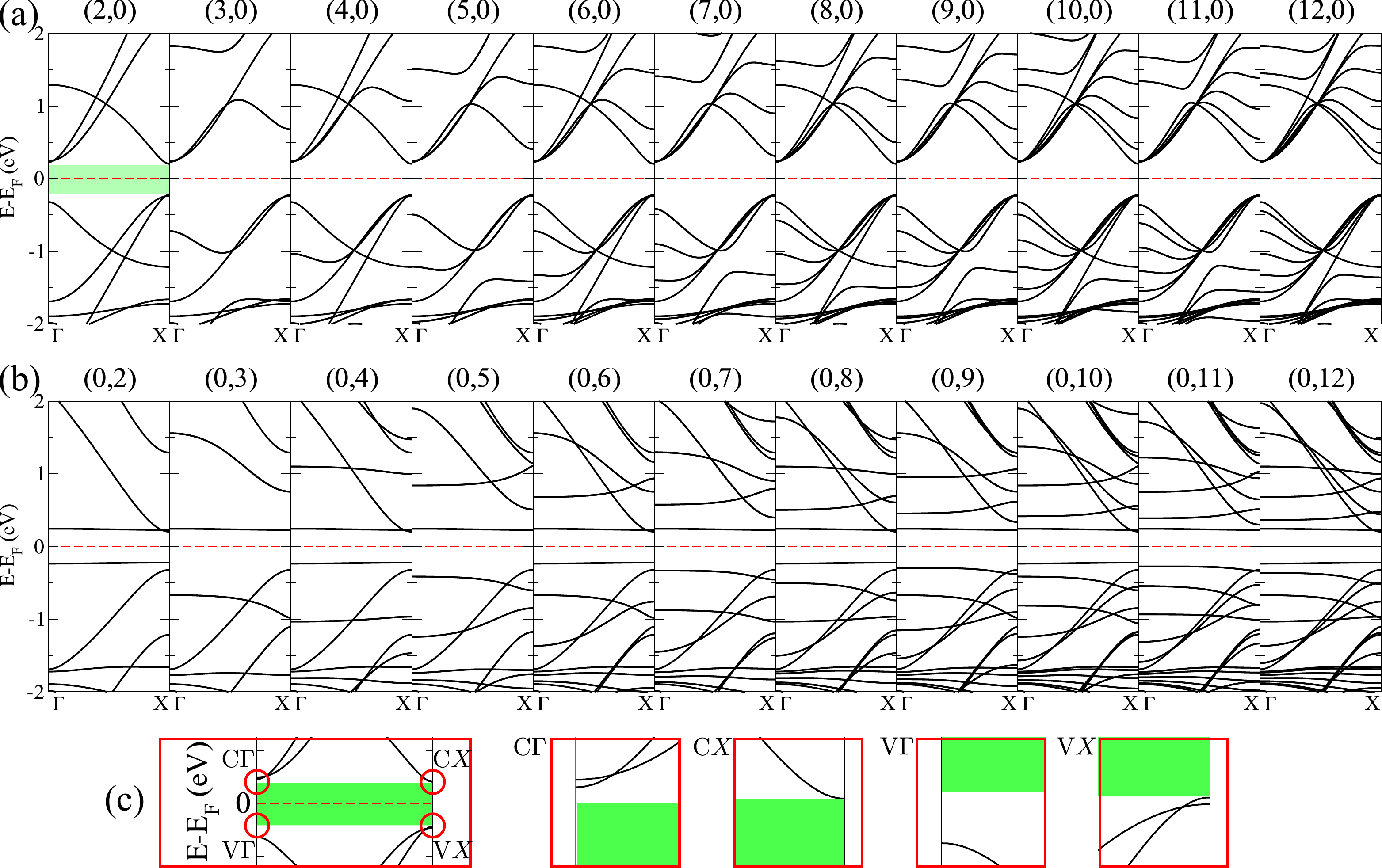}
\caption{(a) Electronic band structures of the $(n,0)$ nanotubes, with $n$ from 2 to 12, according to the zone-folding approach. The dashed red lines at $E=0$ indicate the Fermi level. (b) Same as (a), but for the $(0,m)$ nanotubes, with $m$ from 2 to 12. (c) Insets of the frontier states of the valence and conduction bands of the $(2,0)$ tube, with the gap energy region highlighted in green.}
\label{fig-bands-zf}
\end{figure*}

The cutting lines for a $(n,0)$ armchair nanotube are parallel to the $\Gamma-Y$ and $X-S$ directions of the 2D BZ. The band structure along these paths is highlighted by the cyan shaded areas in the band structure from Fig.~\ref{fig-rggy-ee}(a). In the upper-central part of Fig.~\ref{fig-rggy-ee}(e), we illustrate the cutting lines of a $(4,0)$ nanotube over the r$\gamma$GY BZ. In the left-hand side of Fig.~\ref{fig-rggy-ee}(e), we further show the projection of these $(4,0)$ lines over the band structure of r$\gamma$GY. From a ZF perspective, we expect a set of $N$ ($N=n$ for the $(n,0)$ tube) conduction bands to be almost degenerated at the $\Gamma$ point, as all cutting lines cross the $k$-space line ($\Gamma$ to $X$) of the flat sector of the conduction band. In particular, the $(4,0)$ tube has a feature shared with all the $(n,0)$ nanotubes with even $n$: a cutting line passing along the $X-S$ path (see Fig.~\ref{fig-rggy-ee}(e)). This is a relevant property, since the global minimum of the conduction band lies at the $S$ point, in the dispersive sector of the band. Conversely, odd $n$ nanotubes lack a cutting line at the edge of the 2D BZ (see the $(3,0)$ tube in Fig.~\ref{fig-rggy-ee}(f)). In the ZF picture, the set of $N$ nearly degenerated conduction bands at the $\Gamma$ point is present in all $(n,0)$ tubes, but in the odd-$n$ case, this set usually defines the CBM. On the other hand, in $(n,0)$ armchair nanotubes with even $n$, the CBM lies at the BZ edge ($S$ point in the 2D system, corresponding to the $X$ point in the nanotube BZ) due to the $j=n/2$ cutting line. We note that for odd $n$, there is no cutting line passing through the $X$–$S$ path; however, when $n$ is sufficiently large, the line closest to this path may intersect the dispersive part of the conduction band at a point very close to $S$, at an energy lower than that of the flat portion of the same band. These aspects can be further seen from the ZF-calculated band structures for $(n,0)$ r$\gamma$GYNT shown in Fig.~\ref{fig-bands-zf}(a) for $n$ from 2 to 12. We note that the local minimum of the conduction band at $X$ for odd $n$ tubes moves toward the Fermi energy as we increase tube diameter, but even the widest odd $n$ tube of this series still has the CBM at the $\Gamma$ point. On the other hand, all even $n$ tubes have the CBM at the edge of the BZ, coming from the $j=N/2$ cutting line.

\begin{figure*}[ht!]
\includegraphics[width=\textwidth]{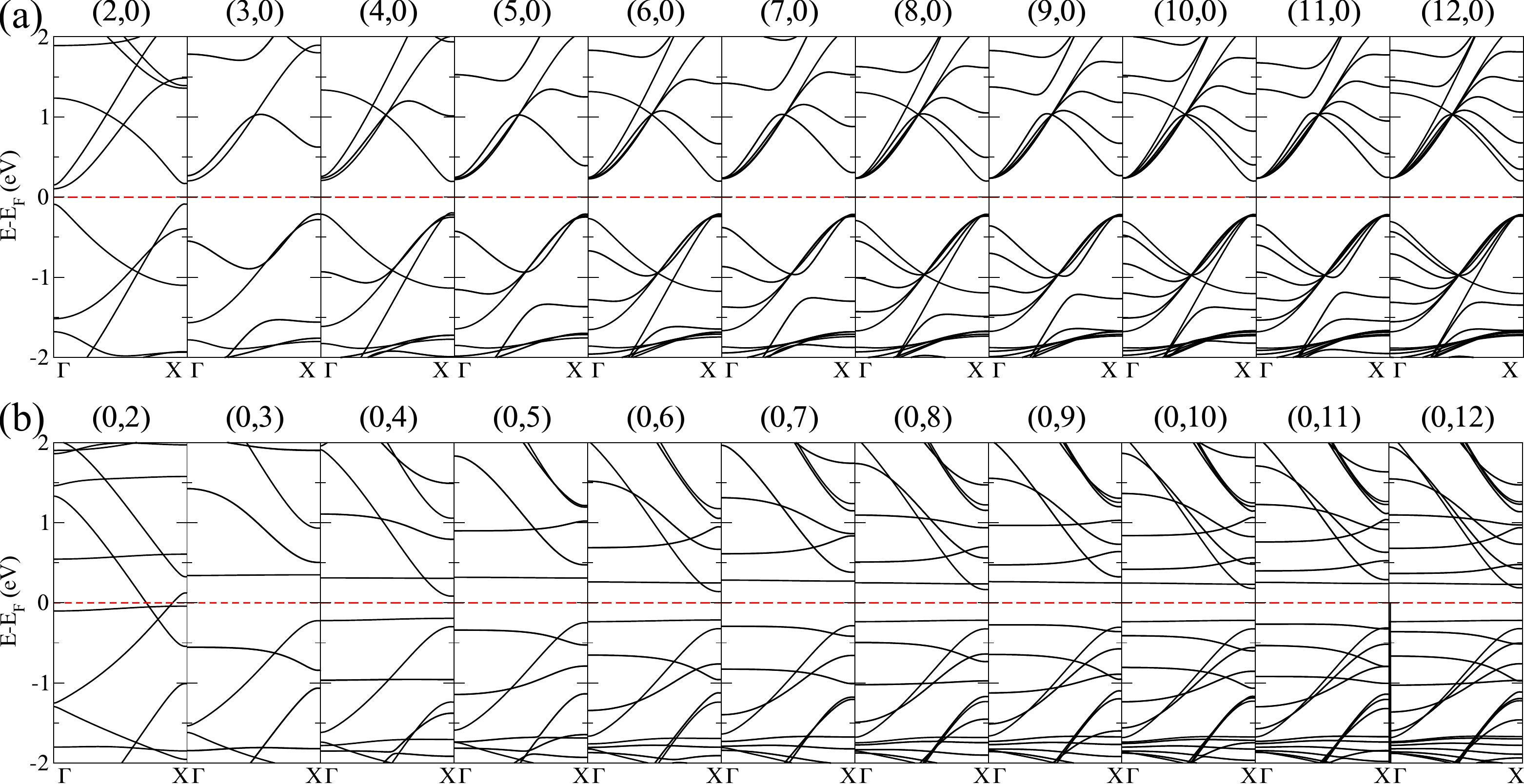}
\caption{(a) Electronic band structures of the $(n,0)$ nanotubes, with $n$ from 2 to 12, according to explicit DFT calculations (including a previous geometry optimization). The dashed red lines at $E=0$ indicate the Fermi level. (b) Same as (a), but for the $(0,m)$ nanotubes, with $m$ from 2 to 12.}
\label{fig-bands-dft}
\end{figure*}

For the valence band, we also have $N$ almost degenerated valence bands, but now at the edge of the tube's BZ ($X$ point of the 1D BZ). Similarly to the conduction band, this occurs because all cutting lines cross the $Y-S$ path, where the flat sector of the valence band lies. This corresponds to the system's VBM for all $n$ values, as per ZF. Also, in the even $n$ cases, the $(n,0)$ tubes will feature a lower valence ZF band with a maximum at $\Gamma$, differently from the odd $n$ cases, as exemplified by the cases $(4,0)$ and $(3,0)$ in Fig.~\ref{fig-rggy-ee}(e-f), respectively. In summary, the ZF approach predicts dispersive valence and conduction bands for armchair r$\gamma$GYNTs, with these bands featuring maxima (for the valence band) and minima (for the conduction band) at both the center and the edge of the 1D BZ. Physically, this dispersive character of the nanotubes bands is compatible with the delocalization of 2D r$\gamma$GY frontier states, which spread like quasi-1D states along the $\mathbf{a}_2$ direction of the sheet (the same direction of the armchair nanotubes axis). From the ZF results in Fig.~\ref{fig-bands-zf} for $(n,0)$ tubes with $n$ from 2 to 12, we can summarize that the CBM is always at $X$ for even $n$, while the conduction band value at $X$ is shifted up for the odd $n$ cases, with the CBM coming from the $\Gamma$ point. These aspects are further highlighted for the $(2,0)$ case in Fig.~\ref{fig-bands-zf}(c), where we amplify the conduction bands at $\Gamma$ and $X$ in the upper close vicinity of the band gap. For the VBM, it always comes from the $X$ point, but the value of the valence band at $\Gamma$ approaches (moves downwards from) the VBM for even (odd) $n$.
This is also highlighted for the $(2,0)$ case in Fig.~\ref{fig-bands-zf}(c), where we amplify the valence bands maxima at $\Gamma$ and $X$ in the lower close vicinity of the band gap. Making a further comparison with the 2D parent system, even $n$ $(n,0)$ nanotubes have their CBM originating from the E3 level of the sheet ($S$ point of the 2D BZ, $X$ point of the 1D BZ), while the CBM from odd $n$ cases arises from E1/E2-like states ($\Gamma-X$ line of the 2D BZ, $\Gamma$ point of the 1D BZ). For the VBM, it always originates from the E5/E6-like 2D levels ($Y-S$ line of the 2D BZ, $X$ point of the 1D BZ), while a lower local valence band maximum originates from E4 ($X$ point of the 2D BZ, $\Gamma$ point of the 1D BZ). In this way, the ZF calculated band gap for even $n$ $(n,0)$ tubes is always the same as that of the 2D counterpart ($\sim$0.42 eV), while it is a little wider for odd $n$ cases ($\sim$0.45 eV at least up to the $n=11$ case).

In the $(0,m)$ cases, the cutting lines are now parallel to the $\Gamma-X$ and $Y-S$ paths. These are the paths along which we observe the flat sectors of the r$\gamma$GY's frontier bands, as highlighted by the yellow shaded areas in the band structure of r$\gamma$GY in Fig.~\ref{fig-rggy-ee}(a). We always expect a flat conduction band for $(0,m)$ tubes according to ZF, since we always have a cutting line along $\Gamma-X$ ($j=0$). However, for even $m$ cases, we also have a cutting line along $Y-S$, and a high-dispersive conduction band crosses the flat band, resulting in the system's CBM close to the nanotube's $X$ point. In the odd $m$ cases, such a crossing line is absent, and the system's CBM typically originates from the flat band (except for huge tubes, which have a high density of cutting lines). Even for the widest odd $m$ system we investigated, the CBM still comes from the flat band. These features can be observed in Fig.~\ref{fig-rggy-ee}(e-f), where we show the cutting lines and the corresponding projection of the 2D bands for the $(0,4)$ and $(0,3)$ nanotubes, respectively. For the valence band, ZF always predicts a dispersive band with its maximum at the BZ's edge, originating from the zero-th cutting line. This band typically contains the VBM in the odd $m$ cases, whereas the VBM originates from another flat band that appears in the even $m$ cases, due to the cutting line passing by $Y-S$. In Fig.~\ref{fig-bands-zf}(b), we show the band structures obtained by ZF for the $(0,m)$ tubes with $m$ from 2 to 12. We observe that there is always a conduction flat band. But the CBM lies at such a non-dispersive band only in the odd $m$ cases, while the minimum of the high-dispersive conduction band at $X$ is the CBM for even $m$, as it crosses the flat branch. For the occupied states, we have a flat valence band (containing the system's VBM) for the even $m$ cases. Such a band is absent for odd $m$ and the maximum value of a dispersive valence band at $X$ is the VBM. Even though the flat band of the cutting line along $Y-S$ is absent in odd $m$, another band, which is flat over most of the 1D BZ, starts to develop for large $m$, as the density of cutting lines increases and we have a line close to (but not at) the $Y-S$ path. Such a partially flat band in odd $m$ cases gradually approaches the dispersive VBM, while they eventually cross each other at (0,9), where the CBM now comes at $\Gamma$ from the low-dispersive band. In summary, the CBM in odd $m$ cases usually comes from E2 states from r$\gamma$GY (which are slightly below E1 levels), while it comes from the E3 state for even $m$. Conversely, the VBM comes from the E5 states (a little higher in energy compared to E6) for even $m$, while it comes from the E4 state for odd $m$ (or from a state close to E5 for large $m$ values). In this way, ZF predicts that $(0,m)$ tubes with even $m$ to have the same gap as the 2D counterpart, while it is a little wider (at most $\sim$0.55 eV) for odd $m$ cases.

We note that ZF yields more accurate results for very large tubes. Although qualitative considerations can be made for narrow tubes based on ZF, it fails to provide precise bands when a high curvature occurs. In this way, especially for small-diameter tubes, it is crucial to proceed with explicit simulations that involve geometry optimization and further computation of the bands for these cylindrical materials. 

\begin{figure}[t!]
\includegraphics[width=\columnwidth]{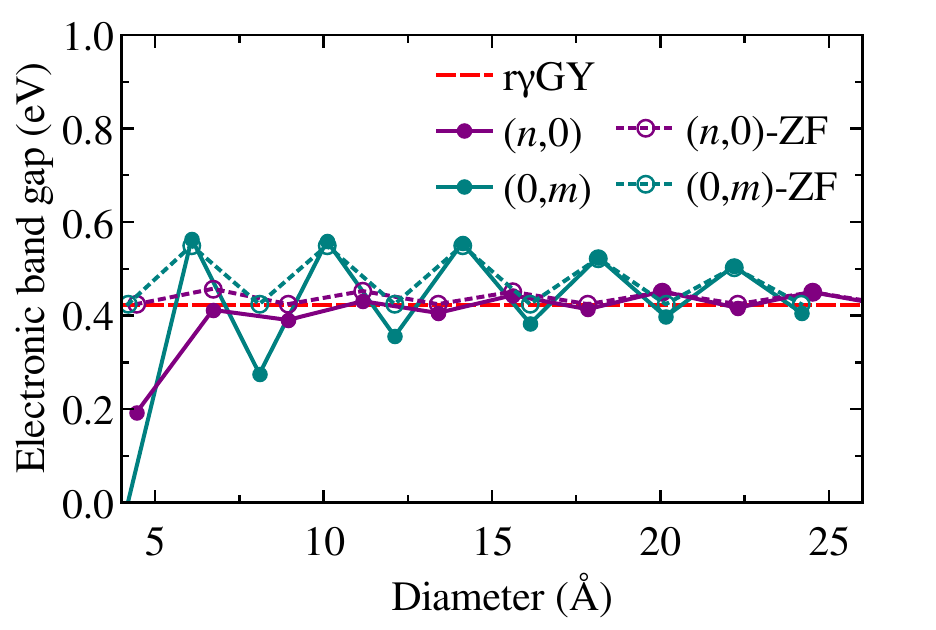}
\caption{Electronic band gap of $(n,0)$ (purple lines) and $(0,m)$ (green lines) r$\gamma$GY nanotubes as a function of tube diameter according to the ZF (empty circles with dashed lines) and DFT (filled circles with full lines) approaches. The horizontal red dashed line indicates the 2D r$\gamma$GY reference value.}
\label{gap-diam}
\end{figure}

In Fig.~\ref{fig-bands-dft}(a), we show the electronic band structure for the $(n,0)$ nanotubes calculated from an explicit DFT calculation. We note that the DFT results agree with the ZF data in qualitative terms, even for narrow tubes. We also observe a reasonable quantitative agreement for wider tubes. In general, the explicit effect of curvature is to move the DFT frontier bands closer to the $E_F$ compared to the ZF picture, resulting in overall smaller gaps. The (2,0) tube is the one featuring the major DFT-ZF differences, even though ZF is still able to predict the semiconducting character of the tube, with (2,0) DFT (ZF) gap being 0.19 eV (0.42 eV). Another DFT-ZF difference in the (2,0) case is that the CBM arrives at the $\Gamma$ point in DFT, instead of at $X$ as expected from ZF. Furthermore, the set of almost degenerate conduction (valence) bands at $\Gamma$ ($X$) is also lifted in the (2,0) tube, a feature also easily noticeable in the (3,0) case, while this effect fades away as the tube diameter increases. In Fig.~\ref{gap-diam} we plot the band gap of the $(n,0)$ tubes as a function of tube diameter according to the DFT and ZF approaches. The DFT results maintain the even-odd oscillations predicted by ZF. We gradually observe a better performance of FZ predicting the band gap, as the even $n$ DFT series shows a gap within an 11 meV difference to that of the 2D counterpart value for $n\ge8$. On the other hand, except for the $n=3$ case, all tubes from the odd series feature a gap larger than the 2D case (a ZF prediction), while the DFT-ZF difference in the band gap value lies below 0.1 eV for $n\ge7$. 

\begin{figure}[t!]
\includegraphics[width=\columnwidth]{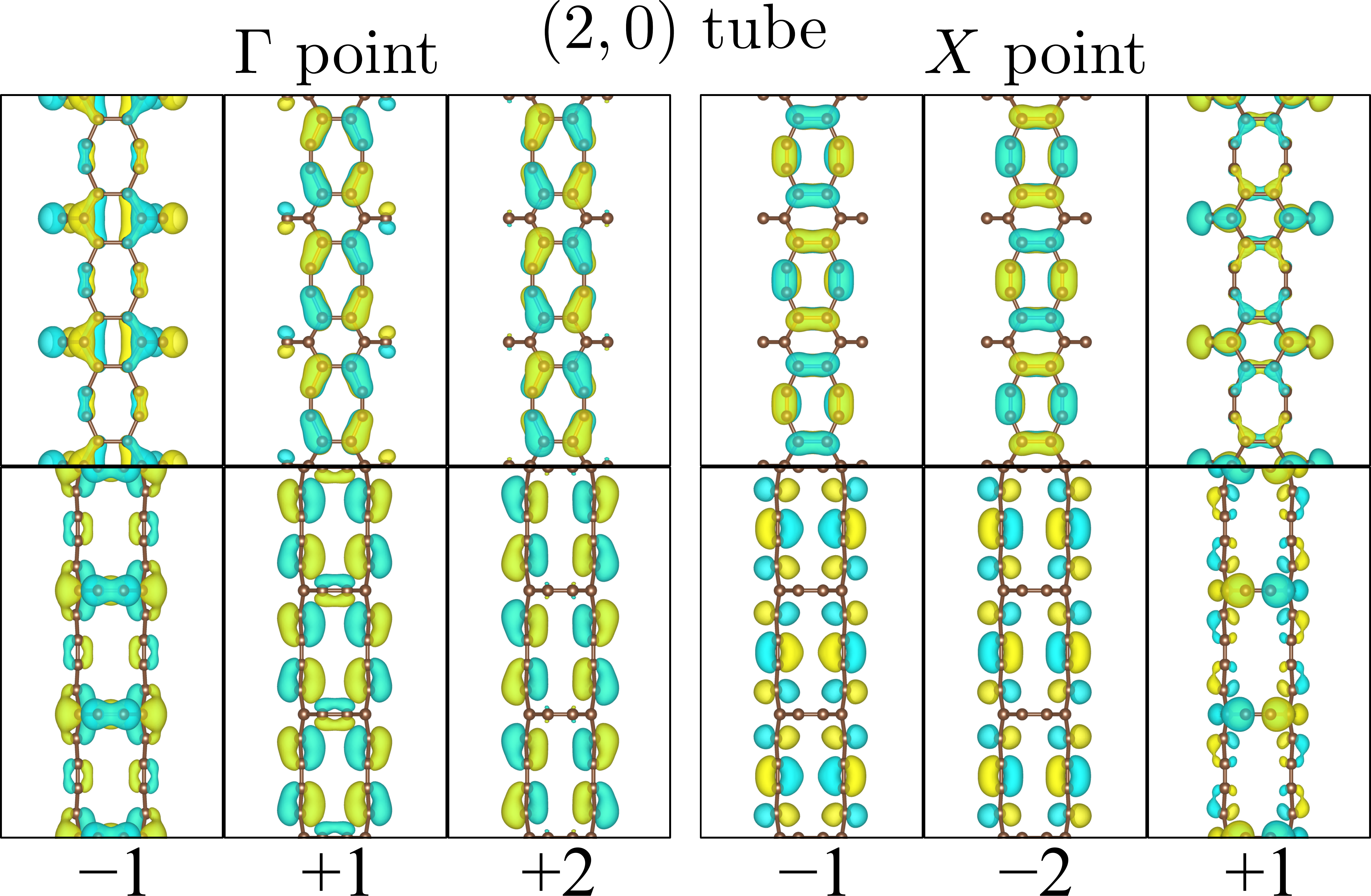}
\caption{Real part of the wavefunctions of the (2,0) tube for the highest valence band ($-1$) and for the two lowest conduction bands ($+1$ and $+2$) at the $\Gamma$ point, as well as for the highest two valence bands ($-2$ and $-1$) and for the lowest conduction band ($+1$) band at the $X$ point of the 1D BZ.}
\label{20-wave}
\end{figure}

In Fig.~\ref{20-wave}, we illustrate particular levels from the $(2,0)$ by plotting two side views of the real part of their wavefunctions. Here, we consider a phase choice in which the imaginary part vanishes. We plot the wavefunctions for the highest valence ($-1$) and for the first two conduction bands ($+1$ and $+2$) at the $\Gamma$ point, as well as for the highest two valence ($-2$ and $-1$) and for the conduction ($+1$) band at the $X$ point of the 1D BZ. We note that the valence band wavefunction at $\Gamma$ and the conduction band wavefunction at $X$ (both in the 1D BZ) hold similarities with the E4 and E3 states in the 2D counterpart, which is consistent with the ZF association between tube and 2D levels. Furthermore, E4/E3 in r$\gamma$GY lies at the $X$/$S$ point of the 2D BZ, while we note that the $\mathbf{b}_1$ component of these wavevectors is 1/2 ($\mathbf{b}_1$ being the r$\gamma$GY reciprocal lattice vector along the $\mathbf{a}_1$ direction of the sheet). In the 2D system, the E4/E3 wavefunctions change sign as we move to neighboring cells along $\mathbf{a}_1$, recovering the original phase for even multiples of such a displacement. This is compatible with the geometry of the $(2,0)$ tube, as it has two r$\gamma$GY unit cells around its circumference. On the other hand, the wavefunctions for the first two conduction bands at $\Gamma$ have very similar features compared to the E1/E2 states from the sheet. While these tube states do not change sign when moving between neighboring cells along the axis (a $\Gamma$ point feature), the $+1$ ($+2$) wavefunction does not change (change) sign for successive r$\gamma$GY cells along the circumference, consistent with the symmetry of their originating $k$-vectors in the 2D system. In the $+1$ case, we note that the atoms from the SABs have a significant contribution resulting from curvature. Such an amplitude on the SABs resembles $p$ orbital lobes orthogonal to the plane containing the C-C bonds, different from the rest of the structure, where the $\pi$-like clouds are somehow orthogonal to the tube surface. Similarly, the spatial distribution of the valence $-1$ and $-2$ states at the tube $X$ point confirms their origin from the E6/E5 of the parent sheet, as well as the $+1$ state compared to E3.

\begin{figure*}[t!]
    \centering
    \includegraphics[width=\linewidth]{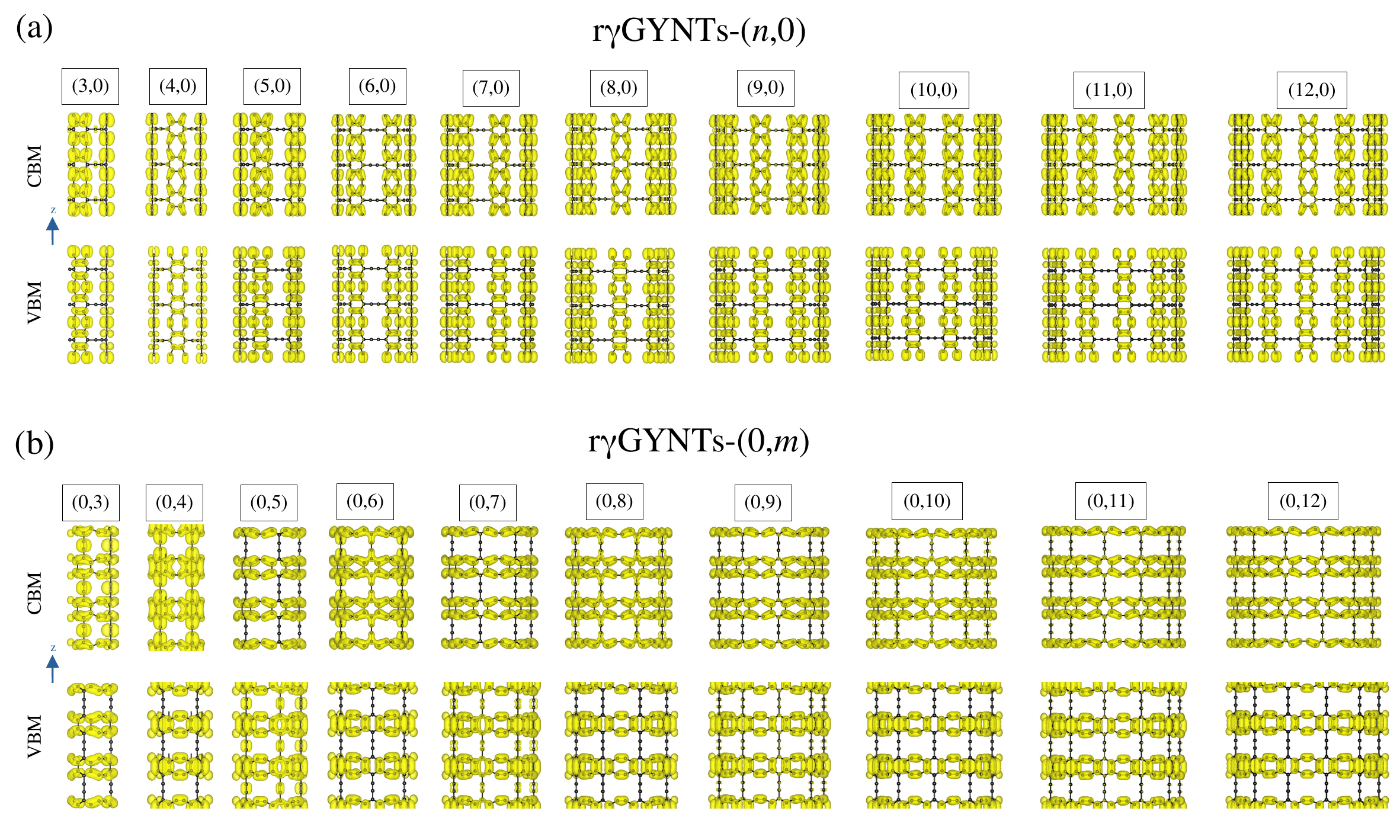}
    \caption{(a) LDOS plots for the VBM and CBM levels of the $(n,0)$ r$\gamma$GYNTs with $n$ from 3 to 12. (b) Same as (a), but for $(0,m)$ r$\gamma$GYNTs with $m$ from 3 to 12.}
    \label{fig:cbm-vbm-tubes}
\end{figure*}

In Fig.~\ref{fig-bands-dft}(b), we show the electronic band structure for the $(0,m)$ nanotubes calculated from an explicit DFT calculation. These results agree well with the ZF data in qualitative terms for most cases. The only exception is the $(0,2)$ tube, which is predicted as being metallic within DFT. Overall, the flat (dispersive) conduction band is shifted away (towards) the Fermi energy, while both flat and dispersive valence levels shift closer to the $E_F$. This results in DFT gaps generally narrower than the corresponding ZF values. Such an effect reaches its maximum strength in the $m=2$ case, resulting in bands that cross the Fermi level. Even in this case, the DFT results still capture the pair of flat bands predicted from ZF. For the even $m$ series, all tubes are predicted to have the same gap as the 2D systems according to ZF, whereas we observe that the DFT-predicted gap approaches that of the 2D r$\gamma$GY at a slower rate compared to the $(n,0)$ nanotubes. We can rationalize this in terms of the positions of the acetylenic chains relative to the axis of the tube. We note that the CABs lie around the tube circumference in $(0,m)$ tubes and are more affected by curvature than the SABs (which lie along the axis). This is the other way around for the $(n,0)$ cases. Since $(0,m)$ tubes have two CABs for each SAB, they are expected to be more affected by curvature than $(n,0)$ systems, which is consistent with the even $n$ cases approaching the gap of the 2D systems at a faster pace than the even $m$ tubes.

To further illustrate the similarity between the ZF and DFT results, in Fig.~\ref{fig:cbm-vbm-tubes} we present the local density of states (LDOS) plots for the VBM and CBM states of the $(n,0)$ and $(0,m)$ r$\gamma$GYNTs, with $n$ and $m$ ranging from 3 to 12. To integrate the LDOS along the energy axis, we considered an interval of 0.01 eV around the VBM/CBM energy value. The results allow us to analyze the electronic distribution of their frontier states along the different structural arrangements. As discussed for the $(n,0)$ tubes, their CBM originates from dispersive (flat) bands of the 2D counterpart for even (odd) $n$. In this sense, we would expect that the CBM of $(n,0)$ tubes would resemble the E3 (E1/E2) states in even (odd) $n$ systems. However, these levels are very close in energy on the 2D parent system and in their 1D counterparts with an even $n$ index, so that both tube levels are summed up once we integrate the LDOS over the 0.01 eV interval. And since conduction (E1/E2-like) states feature a much higher DOS contribution, they dominate the CBM LDOS profile for all $(n,0)$ tubes, as observed in Fig.~\ref{fig:cbm-vbm-tubes}(a). On the other hand, the VBM in the $(n,0)$ tubes always comes from the flat branch of the originating 2D layer, so the VBM for all these tubes looks like the E5/E6 2D states.

For the $(0,m)$ tubes, the even-odd alternation clearly shows up in the LDOS profiles. For even $m$ tubes, their VBM originates from the 2D dispersive valence band, so that the tubes' LDOS has a strong participation of the SABs (see Fig.~\ref{fig:cbm-vbm-tubes}(b)), a feature from the 2D E3 level. For odd cases, the CBM originates from the flat bands, and their LDOS clearly resembles the spatial distribution of the E1/E2 2D states. Regarding the VBM, it always comes from the flat band in even $m$ tubes, and their LDOS plots have a similar profile when compared to the 2D E5/E6 states. This is different for the odd $m$ tubes, as the VBM originates from the dispersive 2D valence band (E4) states. As we look at the $(0,3)$ and $(0,5)$ tubes, there is a strong participation of the SAB atoms (a feature of the 2D E4 states). However, as we increase the tube diameter, the flat sector of the valence band approaches the VBM and gradually becomes closer to the VBM in odd $m$ systems (see Fig.~\ref{fig-bands-dft}(b)), entering the energy range used for LDOS integration. So, starting from the $(0,7)$ system to wider tubes, the relative LDOS contribution from the SAB atoms becomes less noticeable, and we observe a charge profile closer to that of the 2D E5/E6 states, as the flat portion of the tubes' valence bands provides a much larger contribution to the total DOS.

\begin{figure}[t!]
\includegraphics[width=\columnwidth]{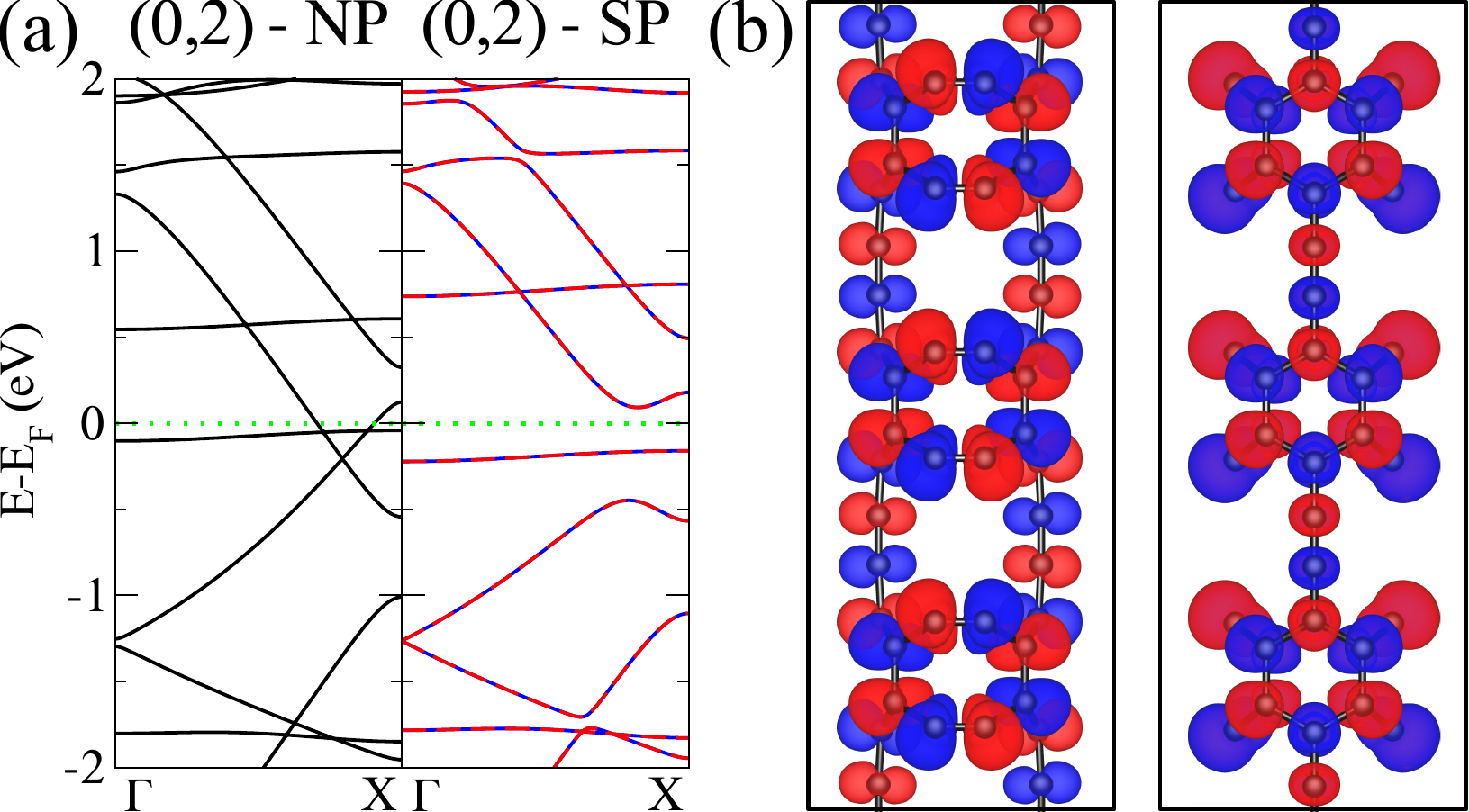}
\caption{(a) Electronic band structure of the (0,2) tube in its NP (black full lines) and SP (blue full/red dashed lines for spin-up/spin-down) configurations. Dotted green lines represent the Fermi energy. (b) Spin-polarization plot for the SP state of the (0,2) tube, with blue/red clouds representing regions with an excess of spin-up/spin-down electrons.}
\label{02-spin}
\end{figure}

We finish discussing another feature of the $(0,2)$ tube: a flat band very close to the Fermi level. In other systems that share this characteristic, it is very common to observe the occurrence of spin-polarized (SP) states~\cite{pisani2007}. In addition, a SP configuration emerges in systems with structural similarity to r$\gamma$GY, such as 2D biphenilene~\cite{alcon2022} and a rectangular $\alpha$-GY-like version of biphenylene~\cite{rego2025}. In the last case, the 2D system has flat levels like the ones shown by r$\gamma$GY, but such levels in the rectangular $\alpha$-GY system lie much closer to $E_F$ than in r$\gamma$GY (which does not feature such a SP configuration). For this reason, we further performed spin-polarized calculations for the $(0,2)$ tube, which revealed that it hosts a non-trivial spin electronic distribution. In Fig.~\ref{02-spin} we show the electronic band structure of the SP state of the $(0,2)$, where blue full (red dashed) lines represent spin-up (spin-down) bands. We also reproduce the band structure of the corresponding non-polarized (NP) state (black full lines, also previously shown in Fig.~\ref{fig-bands-dft}). The SP state is a semiconducting configuration, as it shifts the valence flat band outwards of the Fermi level, and lifts (by 0.54 eV) the crossing degeneracy of the two dispersive bands from the NP case. However, the band gap of the system is between the flat valence band and the upper dispersive branch (0.25 eV). The difference between the spin-up and spin-down components of the electronic charge is further illustrated by the plot in Fig.~\ref{02-spin}, where regions with an excess of spin-up (spin-down) are illustrated by blue (red) color clouds. We note that each atom with a majority spin-up is neighboring to atoms with a majority spin-down, resulting in a configuration with a zero total magnetic moment. The onsite spin orientations are compatible with the bipartition of the atomic structure, similar to what happens with 2D biphenylene and the rectangular $\alpha$-GY based on biphenylene. The SP configuration is also the system's ground state by an 88 meV difference.

\section{Concluding remarks}

In summary, we investigated the electronic properties of carbon nanotubes based on a rectangular graphyne sheet. In general, large-diameter tubes closely follow the properties of their parent sheet, as these tubes are also semiconducting. The overall profiles of the band structures of low-curvature tubes are well described in terms of a zone-folding approach, which rationalizes the resemblance between the electronic signatures of tubes and their originating sheets. However, curvature acts as a way to modulate the tubes' band gap, especially for tubes with the $(0,m)$ chirality, as the folding direction structurally affects mainly acetylenic bridges strongly associated with the frontier states of the original 2D system. Strong gap modulation also occurs for narrow nanotubes, including a case where a nanotube becomes metallic. This last structure undergoes a further transition to a semiconducting state due to its spin-polarized configuration. A special advantage of r$\gamma$GYNTs is that they are systematically semiconductors, regardless of their chirality, a distinct feature compared to graphene-based nanotubes, which are semiconducting only under specific structural conditions.

\section{Acknowledgements}
M.L.P.J. acknowledges financial support from FAPDF (grant 00193-00001807/2023-16), CNPq (grants 444921/2024-9 and 308222/2025-3), and CAPES (grant 88887.005164/2024-00). E.C.G. acknowledges support from CNPq (Process No. 309832/2023-3) and from the Laborat\'orio de Simula\c c\~ao Computacional Caju\'ina (LSCC) at Universidade Federal do Piau\'i. The authors also acknowledge support from the Centro Nacional de Processamento de Alto Desempenho at Ceará (CENAPAD-UFC) and São Paulo (CENAPAD-SP) in Brazil.

\bibliographystyle{edu_style2}
\bibliography{bibs}

\end{document}